\RequirePackage{fix-cm}
\documentclass[smallextended]{svjour3}       
\smartqed  
\usepackage{bm}
\usepackage[dvips]{graphicx}
\begin{document}

\title{Large-scale empirical study on pairs trading for all possible pairs of stocks  
listed on the first section of the Tokyo Stock Exchange}


\titlerunning{Large-scale empirical study on pairs trading}        

\author{Mitsuaki Murota      \and   Jun-ichi Inoue 
}


\institute{Mitsuaki Murota  and Jun-ichi Inoue\at
              Graduate School of Information Science and Technology, 
Hokkaido University, N14-W9, Kita-ku, Sapporo 060-0814, Japan \\
              Tel.: +81-11-7067225\\
              Fax: +81-11-7067391\\
              \email{\{murota,\,j$\underline{\,\,\,}$inoue\}@complex.ist.hokudai.ac.jp, jinoue@cb4.so-net.ne.jp
              }           
}

\date{Received: date / Accepted: date}

\maketitle

\begin{abstract}
We carry out a large-scale empirical data analysis 
to examine the efficiency of the so-called pairs trading. 
On the basis of relevant three thresholds, 
namely, starting, profit-taking,  and stop-loss for the `first-passage process' of the spread (gap) 
between two highly-correlated stocks, we construct an effective strategy 
to make a trade via `active' stock-pairs automatically. 
The algorithm is applied to $1,784$ stocks 
listed on the first section of the Tokyo Stock Exchange leading up to totally $1,590,436$ pairs. 
We are numerically confirmed that the asset management by means of the pairs trading works effectively at least for  
the past three years (2010-2012) 
data sets in the sense that 
the profit rate becomes positive (totally positive arbitrage) in most cases of the possible combinations of thresholds 
corresponding to `absorbing boundaries' in the literature of first-passage processes. 

\end{abstract}
\section{Introduction}
Cross-correlations often provide us very useful information about financial markets to 
figure out various non-trivial and complicated structures behind the stocks as multivariate time series \cite{Bouchaud_book}. 
Actually, 
the use of the cross-correlation 
can visualize collective behavior of stocks during the crisis. 
As such examples, we visualized the collective movement of the stocks 
by means of the so-called multi-dimensional scaling (MDS) 
during the earthquake in Japan on March 2011 \cite{Ibuki,Ibuki2,Ibuki3}. 
We have also constructed a prediction procedure for several stocks 
simultaneously by means of multi-layer Ising model having mutual correlations through 
the mean-fields in each layer \cite{Ibuki,Ibuki2,Ibuki3,Murota}. 

Usually, we need information about the trend of each stock to predict 
the price for, you might say,  `single trading' \cite{Murota,Kaizouji2000,Bouchaud}. 
However, 
it sometimes requires us a lot of unlearnable `craftsperson's techniques' to make a profit. 
Hence, it is reasonable for us to 
use the procedure without any trend-forecasting-type way 
in order to manage the asset with a small risk. 

From the view point of time-series prediction, 
Elliot  {\it et.al.} \cite{Elliot} 
made a model for the spread 
and tried to estimate the state variables (spread) as 
hidden variables from observations by means of Kalman filter. 
They also estimated the hyper-parameters 
appearing in the model by using EM algorithm 
(Expectation and Maximization algorithm) which has been used in 
the field of computer science.   
As an example of constructing optimal pairs, 
Mudchanatongsuk \cite{Mud} 
regarded pair prices as 
Ornstein-Uhlenbeck process,  and 
they proposed a portfolio optimization 
for the pair by means of stochastic control.

For the managing of assets,  
the so-called {\it pairs trading} \cite{Vidy,Mark,Gatev} has attracted trader's attention. 
The pairs trading is based on the assumption that 
the spread between highly-correlated two stocks 
might shrink eventually even if the two prices of the stocks temporally exhibit `mis-pricing' leading up to a large spread.  
It has been believed that the pairs trading is almost `risk-free' procedure, 
however, there are only a few extensive studies \cite{Marcelo,Do} so far to examine the conjecture 
in terms of big-data scientific approach.

Of course, several purely theoretical approaches based on probabilistic theory have been reported. 
For instance, 
the so-called 
{\it arbitrage pricing theory (APT)} \cite{Gatev} in the research field of 
econometrics 
has suggested that 
the pairs trading works effectively if the linear combination of two stocks, each of which is non-stationary time series, 
becomes stationary. Namely, the pair of two stocks showing the properties of the so-called {\it co-integration} \cite{Engle,Stock} might be a suitable pair. 
However, it might cost us a large computational time 
to check the stationarity of the co-integration for all possible pairs in a market,  whereas 
it might be quite relevant issue to clarify whether 
the pairs trading is actually safer than the conventional `single trading' (see for instance \cite{Murota}) to manage the asset, 
or to what extent  the return from the pairs trading would be expected {\it etc}.

With these central issues in mind, 
here we construct a platform to carry out and to investigate  the pairs trading 
which has been recognized an effective procedure for some kind of `risk-hedge' in asset management. 
We propose an effective algorithm (procedure) to check the amount of profit from the pair trading easily and automatically. 
We apply our algorithm to daily data of stocks in the first section of the Tokyo Stock Exchange, 
which is now available at the Yahoo!\,finance web site \cite{Yahoo}.  
In the algorithm, 
three distinct conditions, namely, 
starting ($\theta$),  profit-taking ($\varepsilon$) and stop-loss ($\Omega$) conditions of transaction 
are automatically built-into the system by evaluating the spread (gap) between the prices of two stocks for a given pair. 
Namely, we shall introduce three essential conditions to inform us when we should start the trading, 
when the spread between the stock prices satisfies the profit-taking conditions, {\it etc}. by 
making use of a very simple way. 
Numerical evaluations of the algorithm for the empirical data set 
are carried out for all possible pairs 
by changing the starting, profit-taking and stop-loss conditions 
in order to look for the best possible combination of the conditions. 

This paper is organized as follows. 
In the next section \ref{sec:sec2}, 
we introduce several descriptions for the mathematical modeling of pairs trading and 
set-up for the empirical data analysis by defining various 
variables and quantities. 
Here we also mention that the pairs trading is described by a first-passage process \cite{Redner},  
and explain the difference between our study and arbitrage pricing theory (APT) \cite{Gatev} which have highly developed in the research field of econometrics. 
In section \ref{sec:sec3}, 
we introduce several rules of the game for the trading. 
We define two relevant measurements to quantify the usefulness of pairs trading, 
namely, winning probability and profit rate. 
The concrete algorithm to carry out pairs trading automatically is also given in this section explicitly. 
The results of empirical data analysis are reported and argued in section \ref{sec:sec4}. 
The last section is devoted to summary. 
\section{Mathematical descriptions and set-up}
\label{sec:sec2}
In pairs trading, we first pick up two stocks having a large correlation in the past. 
A well-known historical example is the pair of {\it coca cola} and {\it pepsi cola} \cite{Pairs}. 
Then, we start the action when the spread (gap) between the two stocks' prices increases up to  
some amount of the level (say, $\theta$), namely, we sell one increasing stock (say, the stock $i$) 
and buy another decreasing one (say, the stock $j$) at the time $t_{<}^{(ij)}$. 
We might obtain the arbitrage as a profit (gain) $g_{ij}$: 
\begin{equation}
g_{ij} = |\gamma_{i}(t_{<}^{(ij)})-\gamma_{j}(t_{<}^{(ij)})|- |\gamma_{i}(t_{>}^{(ij)})-\gamma_{j}(t_{>}^{(ij)})|
\end{equation}
when the spread decreases to some amount of the level (say, $\varepsilon (> \theta)$) again due to 
the strong correlation between the stocks, 
and we buy the stock $i$ and sell the stock $j$ at time $t_{>}^{(ij)}$. 
We should keep in mind that 
we used here the stock price normalized by the value itself at $\tau$-times before 
(we may say `rate') as
\begin{equation}
\gamma_i(t) \equiv 
\frac{p_i (t)-p_i (t-\tau+1)}{p_i (t-\tau+1)} = 
 \frac{p_i(t)}{p_i(t-\tau+1)}-1 
\label{eq:def_gamma}
\end{equation}
where we defined $p_{i}(t)$ as a price of the stock $i$ at time $t$. 
It is convenient for us to use the $\gamma_{i}(t)$ (or $\gamma_{i}(t)+1$) 
instead of the price $p_{i}(t)$ because 
we should treat the pairs changing in quite different ranges of price.  
Hence, we evaluate the spread between two stocks 
by means of the rate $\gamma_{i}(t)$ which 
denotes  how much percentage of the price increases (or decreases) 
from the value itself at $\tau$-times before. 
In our simulation, we choose $\tau=\mbox{250 [days]}$. 
By using this treatment (\ref{eq:def_gamma}), one can use unified thresholds 
$(\theta, \varepsilon, \Omega)$ which are independent of 
the range of prices for all possible pairs.  
\subsection{Pairs trading as a first-passage process}
Obviously, more relevant quantities are now 
not the prices themselves but the 
spreads for the prices of pairs. 
It  might be helpful for us to notice that 
the process of the spread defined by 
\begin{equation}
d_{ij} (t) \equiv  |\gamma_{i}(t)-\gamma_{j}(t)|
\end{equation}
also produces a time series 
$d_{ij}(0) \to 
d_{ij}(1) \to \cdots \to d_{ij}(t) \to \cdots$, 
which 
is described as a stochastic process. 

In financial markets, the spread (in particular, the Bid-Ask spread) is one of the key quantities 
for double-auction systems (for instance, see \cite{Ibuki5}) and 
the spread between two stocks 
also plays an important role in pairs trading. 
Especially, it should be regarded as a {\it first-passage process} (or sometimes referred to as {\it first-exit process}) 
(see for instance \cite{Redner,Inoue,Sazuka,Inoue2,Sazuka2} for recent several applications to finance) 
with absorbing boundaries $(\theta, \varepsilon, \Omega)$, 
and the times $t_{<}^{(ij)}$ and $t_{>}^{(ij)}$ are regarded as 
first-passage times. 
Actually, $t_{<}^{(ij)}$ and $t_{>}^{(ij)}$ 
are the times $t$ satisfying the following for the first time
\begin{equation}
d_{ij} (t) \geq  \theta,\,\,\,\,\mbox{and}\,\,\,\,\, 
\theta < d_{ij} (t)  <  \varepsilon \,\,\,( t<t_{>}^{(ij)}), 
\end{equation}
respectively. 
More explicitly, these times are given by 
\begin{eqnarray}
t_{<}^{(ij)} &  = &  
\min \{t>0 \,|\,  d_{ij}(t) \geq \theta\} \\
t_{>}^{(ij)} & = & 
\min \{t>t^{(ij)}_{<}\, |\, \theta <d_{ij}(t) \leq \varepsilon \}. 
\end{eqnarray}
The above argument was given for somewhat an ideal case, 
and of course, we might lose the money just as much as 
\begin{equation}
l_{ij} =  |\gamma_{i}(t_{*}^{(ij)})-\gamma_{j}(t_{*}^{(ij)})|- |\gamma_{i}(t_{<}^{(ij)})-\gamma_{j}(t_{<}^{(ij)})| 
=|d_{ij}(t_{*}^{(ij)})-d_{ij}(t_{<}^{(ij)})|
\end{equation}
where $t_{*} (>t_{<})$ is a `termination time' satisfying 
\begin{equation}
t_{*}^{(ij)} = \min\{ t >t_{<}^{(ij)} \,|\, d_{ij}(t) > \Omega\}. 
\end{equation}
This means that we should decide a `loss-cutting' when 
the spread $d_{ij} (t)$ does not shrink to the level $\varepsilon$ and 
increases beyond the threshold $\Omega$ at time $t_{*}$. 
It should bear in mind that 
once we start the trading, we get a gain or lose, hence 
the above $t_{>}^{(ij)}$ and $t_{*}^{(ij)}$ are unified as `time for decision' $\hat{t}^{(ij)}$ by 
\begin{equation}
\hat{t}^{(ij)}=
t_{*}^{(ij)} + 
(t_{>}^{(ij)}-t_{*}^{(ij)})
\Theta (t_{*}^{(ij)}-t_{>}^{(ij)})
\label{eq:decision_t}
\end{equation}
where we defined a unit step function as 
\begin{equation}
\Theta (x) = 
\left\{
\begin{array}{cc}
1 & (x\geq 0) \\
0 & (x<0). 
\end{array}
\right.
\end{equation}
Namely, 
if a stochastic process $d_{ij}(t)$ firstly reaches 
the threshold $\varepsilon$, 
the time for decision is $\hat{t}^{(ij)}=t_{>}^{(ij)}$, 
whereas if the $d_{ij}(t)$ goes 
beyond the threshold $\Omega$ before shrinking to the level $\varepsilon$, 
we have $\hat{t}^{(ij)}=t_{*}^{(ij)}$. 
\subsection{Correlation coefficient and volatility}
We already mentioned that the pairs trading is based on the assumption that 
the spread between highly-correlated two stocks 
might shrink shortly even if the two prices of the stocks exhibit a temporal spread.  
Taking into account the assumption, 
here we select the suitable pairs of stocks using the information about 
correlation coefficient (the Pearson estimator) for pairs in order to quantify the correlation:  
\begin{equation}
\rho_{ij} (t)= \frac{\sum_{\Delta t=t-\tau+1}^{t}
(\Delta \gamma_{i} (t, \Delta t)-\overline{\Delta \gamma_{i}(t, \Delta t)})
(\Delta \gamma_{j} (t, \Delta t)-\overline{\Delta \gamma_{j}(t, \Delta t)})}{
\sqrt{
\sum_{\Delta t=t-\tau+1}^{t}(\Delta \gamma_{i} (t, \Delta t)-\overline{\Delta \gamma_{i}(t, \Delta t)})^{2}
\sum_{\Delta t=t-\tau+1}^{t}(\Delta \gamma_{j} (t, \Delta t)-\overline{\Delta \gamma_{j}(t, \Delta t)})^{2}
}}
\label{eq:Pearson}
\end{equation}
and standard deviation 
(volatility): 
\begin{equation}
\sigma_{i} (t) 
=  \sqrt{
\sum_{l=t-\tau+1}^{t} (\gamma_{i}(l)-\overline{\gamma_{i}(t)})^{2}}
\end{equation}
It should be noted that 
we also use the definition of the logarithmic return of the rescaled price 
$\gamma_{i}(t)+1=p_{i}(t)/p_{i}(t-\tau+1)$ (see equation (\ref{eq:def_gamma})) for the duration $\Delta t$ 
in (\ref{eq:Pearson}) 
by 
\begin{equation}
\Delta \gamma_{i}(t, \Delta t) \equiv \log (\gamma_{i}(t+\Delta t)+1)-\log (\gamma_{i}(t)+1)
\label{eq:def_Delta_gamma}
\end{equation}
and moving average of the observable $A(t)$ over the duration $\tau$ as 
\begin{equation}
\overline{A (t)} = 
\frac{1}{\tau}
\sum_{l=t-\tau+1}^{t}
A(l). 
\end{equation}
At first glance, 
the definition of 
(\ref{eq:Pearson}) for correlation coefficient 
might look like unusual because 
(\ref{eq:def_Delta_gamma}) accompanying with (\ref{eq:def_gamma}) implies that 
the correlation coefficient consists of 
the second derivative of prices. 
However, as we already mentioned, 
the `duration' $\tau=1 \mbox{[year]}$ appearing in 
(\ref{eq:def_gamma}) is quite longer than 
$\Delta t =\mbox{1 [day]}$, 
namely, 
$\tau \gg \Delta t$ is satisfied. 
Hence, the price difference in (\ref{eq:def_gamma}) 
could not regarded as the same derivative as in the derivative definition of 
(\ref{eq:def_Delta_gamma}). 
Therefore, the definition of correlation coefficient (\ref{eq:Pearson}) is nothing but 
the conventional first derivative quantity. 

From the view point of these quantities $\{\rho_{ij}(t), \sigma_{i}(t)\}$, 
the possible pairs should be highly correlated and the standard deviation 
of each stock should take the value lying in some finite range. 
Namely, 
we impose the following condition for the candidates of the pairs at time 
$t_{<}^{(ij)}$, that is, 
$\rho_{ij} (t_{<}^{(ij)})  >   \rho_{0}$ and 
$\sigma_{\rm min} < \sigma_{i}(t_{<}^{(ij)}) < \sigma_{\rm max}$. 
Thus, the total number of pairs for us to carry out  the pair trading (from now on, we call such pairs 
as `active pairs') is given explicitly as 
\begin{eqnarray}
N_{(\theta,\varepsilon,\Omega)} & = &  \sum_{i}\sum_{j<i}
\Theta (\rho_{ij}(t_{<}^{(ij)})-\rho_{0})\{
\Theta (\sigma_{i}(t_{<}^{(ij)}) -\sigma_{\rm min})-
\Theta (\sigma_{i}(t_{<}^{(ij)}) - \sigma_{\rm max})
\} \nonumber \\
\mbox{} & \times & 
\Theta (\tau_{\rm max}-\hat{t}^{(ij)})
\label{eq:def_N}
\end{eqnarray}
where a factor $\Theta (\tau_{\rm max}-\hat{t}^{(ij)})$ 
means that we terminate the game 
if `time of decision' $\hat{t}^{(ij)}$ (see equation (\ref{eq:decision_t})) 
becomes longer than the whole playing time 
$\tau_{\rm max}=\mbox{1 [year]}$. 
Therefore, 
the number of active pairs is dependent on the thresholds 
$(\theta,\epsilon, \Omega)$, and 
we see the details of the dependence in Table \ref{tab:tb1} and Table \ref{tab:tb2} 
under the condition $\Omega=2\theta-\varepsilon$ 
(see $N_{w}+N_{l} \equiv N_{(\theta,\varepsilon)}$ in the tables). 

From Fig. \ref{fig:cor}, 
we are also confirmed that 
the number pairs $N$ 
satisfying 
$\rho_{ij} (t_{<}^{(ij)})  >   \rho_{0}$ and 
$\sigma_{\rm min} < \sigma_{i}(t_{<}^{(ij)}) < \sigma_{\rm max}$
is extremely smaller ($N \sim 300$) than the number of combinations for all possible $n$ stocks, 
namely, $N \ll n(n-1)/2=1,590,436$.  
\subsection{Minimal portfolio and APT}
It might be helpful for us to notice that 
the pairs trading could be regarded as a `minimal portfolio' and it  
can obtain the profit even for the case that 
the stock average decreases. 
Actually, it is possible for us to construct such `market neutral portfolio'  \cite{Livan} as follows. 
Let us consider the return of the two stocks $i$ and $j$ which are described as 
\begin{eqnarray}
\Delta \gamma_{i}(t) & = & \beta_{i} \Delta \gamma_{m}(t) +q_{i}(t) 
\label{eq:betai}\\
\Delta \gamma_{j}(t) & = & \beta_{j} \Delta \gamma_{m}(t) +q_{j}(t)
\label{eq:betaj}
\end{eqnarray}
where parameters $\beta_{i},\beta_{j}$ denote the so-called `market betas' 
for the stocks $i,j$, and $\Delta \gamma_{m}(t)$ stands for the return of the stock average, 
namely, 
\begin{equation}
\beta_{i} = \frac{\overline{
\Delta \gamma_{m}(t) \Delta \gamma_{i}(t)}}
{\sqrt{(\overline{\Delta \gamma_{m}(t)})^{2}}},\,\,\,
\beta_{j} =  \frac{\overline{\Delta \gamma_{m}(t) \Delta \gamma_{j}(t)}}
{\sqrt{(\overline{\Delta \gamma_{m}(t)})^{2}}}
\end{equation}
and here we select them ($\beta_{i},\beta_{j}$) as positive values for simplicity. 
On the other hand, $q_{i}(t),q_{j}(t)$ appearing in (\ref{eq:betai})(\ref{eq:betaj}) 
are residual parts (without any correlation with the stock average) 
of the returns of stocks $i,j$. 
Then, let us assume that we take a short position (`selling' in future) of the stock $j$ by volume $r$ and 
a long position (`buying' in future) of the stock $i$. 
For this action, we have the return of the portfolio as 
\begin{equation}
\Delta \gamma_{ij}(t) = 
\Delta \gamma_{i}(t) -r \Delta \gamma_{j}(t) = 
(\beta_{i}-r\beta_{j})\Delta \gamma_{m}(t) + q_{i}(t) -r q_{j}(t). 
\end{equation}
Hence, obviously, the choice of the volume $r$ as  
\begin{equation}
r = \frac{\beta_{i}}{\beta_{j}}
\end{equation}
leads to 
\begin{equation}
\Delta \gamma_{ij}(t) = q_{i}(t) - \left(\frac{\beta_{i}}{\beta_{j}}\right) 
q_{j}(t)
\end{equation}
which is independent of the market (the average stock $\Delta \gamma_{m}(t)$). 
We should notice that 
$\Delta \gamma_{ij}(t) = 
\Delta \gamma_{i}(t) -r \Delta \gamma_{j}(t)$ is rewritten in terms of the profit as follows. 
\begin{eqnarray}
\Delta \gamma_{ij}(t) & = &  
\Delta \gamma_{i}(t) -r \Delta \gamma_{j}(t) = 
\{\gamma_{i}(t+1)-r \gamma_{j}(t+1)\}-
\{\gamma_{i}(t)-r \gamma_{j}(t)\}  \nonumber \\
\mbox{} & \simeq & d_{ij}(t+1)-d_{ij}(t)
\end{eqnarray}
Therefore, in this sense, 
the profit $\Delta \gamma_{ij}(t)$ 
is also independent of the market $\Delta \gamma_{m}(t)$. This empirical fact might tell us the usefulness of paris trading. 

In the arbitrage pricing theory (APT) \cite{Vidy,Gatev}, 
the condition for searching suitable pairs is 
the linear combination of `non-stationary' time series $\gamma_{i}(t)$ and $\gamma_{j}(t)$, 
$\gamma_{i}(t)-r \gamma_{j}(t)$ becomes 
co-integration, namely, 
it becomes `stationary'. 
Then, the quantity possesses 
the long-time equilibrium value $\mu$ and 
we write 
\begin{eqnarray}
\gamma_{i}(t)-r \gamma_{j}(t) & = & \mu +\omega \\
\gamma_{i}(t +l)-r \gamma_{j}(t+l) & = &  \mu - \omega 
\end{eqnarray}
with a small deviation $\omega (>0)$  from the mean $\mu$. 
Therefore, we easily find 
\begin{equation}
\gamma_{i}(t)-r \gamma_{j}(t) - 
\{
\gamma_{i}(t +l)-r \gamma_{j}(t+l)
\} \simeq d_{ij}(t)-d_{ij}(t+l) =2\omega
\end{equation}
namely, we obtain the profit with a very small risk. 
Hence, the numerical checking for the stationarity of 
the linear combination $\gamma_{i}(t)-r \gamma_{j}(t)$ 
by means of, for instance, exponentially fast decay of the auto-correlation function or various types of statistical test 
might be useful for us to select the possible pairs. 
However, it computationally cost us heavily for large-scale empirical data analysis. 
This is a reason why 
here we use the correlation coefficients and volatilities to investigate 
the active pairs instead of the co-integration based analysis as given in 
the references \cite{Vidy,Mark,Gatev,Engle,Stock}. 
\section{Procedures of empirical analysis and `rules of the game'}
\label{sec:sec3}
In this section, 
we explain rules of our game (trading) using 
the data set for the past three years 2010-2012 including 
2009 to evaluate the quantities like correlation coefficient and volatility in 2010 by choosing 
$\tau=250$ [days].  
In following, we explain how one evaluates the performance of pairs trading
according to the rules. 
\subsection{A constraint for the thresholds}
\label{sec:sub_constraint}
Obviously, the ability of the asset management by pairs treading 
is dependent on the choice of the thresholds $(\theta, \varepsilon, \Omega)$. 
Hence, we should investigate how much percentage of total active pairs can obtain a profit for a given set of 
$(\theta, \varepsilon, \Omega)$. 
To carry out the empirical analysis, 
we define the ratio between the profit  $d_{ij}(t_{<}^{(ij)})-d_{ij}(t_{>}^{(ij)}) (>0)$ and 
the loss $d_{ij}(t_{*}^{(ij)})-d_{ij}(t_{<}^{(ij)}) (>0)$ for the marginal spread, 
namely, 
$d_{ij}(t_{<}^{(ij)})=\theta, d_{ij}(t_{>}^{(ij)})=\varepsilon, d_{ij}(t_{*}^{(ij)})=\Omega$ as 
\begin{equation}
\frac{\Omega-\theta}{\theta-\varepsilon} \equiv \alpha
\label{eq:def_alpha}
\end{equation}
where $\alpha (>0)$ is a control parameter. 
It should be noted that for positive constants $\delta,\delta^{'}$ the gap of the spreads (profit)  
$d_{ij}(t_{<}^{(ij)})-d_{ij}(t_{>}^{(ij)})$ is written as 
\begin{equation}
d_{ij}(t_{<}^{(ij)})-d_{ij}(t_{>}^{(ij)}) = \theta +\delta - (\varepsilon -\delta^{'})=
\theta -\varepsilon +(\delta+\delta^{'}) \geq \theta -\varepsilon. 
\label{eq:lower_bound}
\end{equation}
Therefore, 
the difference $\theta -\varepsilon$ appearing in the denominator of 
equation (\ref{eq:def_alpha}) gives a lower bound of the profit. 
Although the numerator $\Omega-\theta$ in (\ref{eq:def_alpha}) has no such an explicit meaning,
however, implicitly it might be regarded as a `typical loss' because the actually realized loss 
$d_{ij}(t_{*}^{(ij)})-d_{ij}(t_{<}^{(ij)})$ fluctuates around the typical value and it is more likely to take 
a value which is close to $\Omega-\theta$. 

Hence, for $\alpha >1$, 
the loss for the marginal spread $\Omega-\theta$ is larger than the lowest possible profit once a transaction is taken place 
and vice versa  for $\alpha <1$. 
If we set $\alpha <1$, it is more more likely to lose the money less than the lowest bound of the profit $\theta -\varepsilon$, 
however, at the same time, it means that we easily lose due to the small gap between $\Omega$ and $\theta$. 
In other words, 
we might frequently lose with a small amount of losses. 
On the other hand, 
if we set $\alpha >1$, 
we might hardly lose, however, once we lose, 
the total amount of the losses is quite large. 
Basically, it lies with traders to decide which to choose 
$\alpha >1$ or $\alpha <1$, 
however, here we set the marginal $\alpha=1$ as a `neutral strategy', that is 
\begin{equation}
\Omega=2\theta-\varepsilon. 
\label{eq:balance2}
\end{equation}
Thus, we have now only two thresholds $(\theta, \varepsilon)$ for 
our pairs trading, and the $\Omega$ should be determined as a `slave variable' from equation
(\ref{eq:balance2}). 
Actually this constraint (\ref{eq:balance2}) can reduce our computational time to a numerically tractable revel. 
Under the condition (\ref{eq:balance2}), we sweep 
the thresholds 
$\theta, \varepsilon$ as 
$0.01\leq \theta \leq 0.09$, $0.0\leq \varepsilon \leq \theta$ ($d\theta =0.01$) and 
$0.1\leq \theta \leq 1.0$, $0.0\leq\varepsilon\leq\theta$ ($d\theta =0.1$) in 
our numerical calculations (see Table \ref{tab:tb1} and Table \ref{tab:tb2}). 
\subsection{Observables}
In order to investigate the performance of pairs trading quantitatively, 
we should observe several relevant performance measurements. 
As such observables, here we define the following wining probability as a function of the thresholds: 
\begin{equation}
p_{w} (\theta,\varepsilon) =  
\frac{
\sum_{i,j<i} \Theta (d_{ij}(t_{<}^{(ij)})-d_{ij}(\hat{t}^{(ij)}))
\psi (d_{ij}(t_{<}^{(ij)}), \rho_{0},\sigma_{\rm min},\sigma_{\rm max}: \theta,\varepsilon)}
{\sum_{i,j<i}
\psi (d_{ij}(t_{<}^{(ij)}), \rho_{0},\sigma_{\rm min},\sigma_{\rm max}: \theta, \varepsilon)}
\end{equation}
where we defined 
\begin{eqnarray}
&& \psi (d_{ij}(t_{<}^{(ij)}), \rho_{0},\sigma_{\rm min},\sigma_{\rm max}: \theta,\varepsilon) \nonumber \\
& \equiv  & 
\Theta (\rho_{ij}(t_{<}^{(ij)})-\rho_{0})\{
\Theta (\sigma_{i}(t_{<}^{(ij)}) -\sigma_{\rm min})-
\Theta (\sigma_{i}(t_{<}^{(ij)}) - \sigma_{\rm max})
\} \Theta (\tau_{\rm max}-\hat{t}^{(ij)}) \nonumber \\
\mbox{} & = & \ll
\Theta (d_{ij}(t_{<}^{(ij)})-d_{ij}(\hat{t}^{(ij)})) \gg \nonumber \\
\mbox{} & =  & \frac{N_{w}}{N_{(\theta,\varepsilon)}} = 1-\frac{N_{l}}{N_{(\theta,\varepsilon)}}
\label{eq:pw}
\end{eqnarray}
where $N_{w},N_{l}$ are numbers of wins and loses, respectively, and 
the conservation of the number of total active pairs 
\begin{equation}
N_{(\theta,\varepsilon)}=N_{w}+N_{l}
\end{equation}
should hold (see the definition of
 $N_{(\theta,\varepsilon,\Omega)}$ in (\ref{eq:def_N}) under 
 the condition $\Omega=2\theta-\varepsilon$). 
The bracket $\ll \cdots \gg$ appearing in 
(\ref{eq:pw}) is defined by 
\begin{equation}
\ll \cdots \gg \equiv 
\frac{
 \sum_{i,j<i}
(\cdots)  
 \psi (d_{ij}(t_{<}^{(ij)}), \rho_{0},\sigma_{\rm min},\sigma_{\rm max}: \theta,\varepsilon) 
}
{
 \sum_{i,j<i}
  \psi (d_{ij}(t_{<}^{(ij)}), \rho_{0},\sigma_{\rm min},\sigma_{\rm max}: \theta, \varepsilon) 
}. 
\end{equation}
We also define 
the profit rate: 
\begin{equation}
\eta (\theta,\varepsilon) = 
\ll d_{ij}(t_{<}^{(ij)})-
d_{ij}(\hat{t}^{(ij)})
\gg
\label{eq:profit}
\end{equation}
which is a slightly different measurement from the winning probability $p_{w}$. 
We should notice that 
we now consider the case with the constraint 
(\ref{eq:balance2}) 
and in this sense,  
the explicit dependences of $p_{w}$ and $\eta$ on $\Omega$ are omitted in the above descriptions. 
We also keep in mind that  
$d_{ij}(t_{<}^{(ij)})-
d_{ij}(\hat{t}^{(ij)})$ takes a positive value if we make up accounts  
for taking the arbitrage at $\hat{t}^{(ij)}=t_{>}^{(ij)}$. 
On the other hand, the $d_{ij}(t_{<}^{(ij)})-
d_{ij}(\hat{t}^{(ij)})$ becomes negative if we terminate the trading due to  loss-cutting. 
Therefore, the above $\eta$ denotes 
a total profit for a given set of the thresholds 
$(\theta,\varepsilon)$. 
\subsection{Algorithm}
We shall list the concrete algorithm for our empirical study on the pairs trading as follows. 
\begin{enumerate}
\item We collect a pair of stocks $(i,j)$ from daily data for the past one year. 
\item Do the following procedures from $t=0$ to \\
$t=\tau (=250 \mbox{: the number of daily data for one year})$. 
\begin{enumerate}
\item  Calculate $\rho_{ij}(t)$ and $\sigma_{i}(t), \sigma_{j}(t)$ to determine whether the pair $(i,j)$ satisfy the start condition. \\
{\bf Start condition: }
\begin{itemize}
\item  If $\sigma_{\rm min} < \sigma_{i}(t), \sigma_{j}(t) < \sigma_{\rm max}$ and 
$\rho_{ij}(t) > \rho_{0}$ and 
$d(t_{<}^{(ij)}) > \theta$, go to (c). 
\item If not, go to (b). 
\end{itemize}
\item $t \leftarrow t+1$ and back to (a). 
\item $t \leftarrow t+1$ and go to the termination condition.  \\
{\bf Termination condition:}
\begin{itemize}
\item If $d_{ij}(t_{>}^{(ij)})<\varepsilon$ (we `win'), 
go to the next pairs $(k,l) \neq (i,j)$. 
\item If not, go back to (c). 
If $d_{ij}(t_{*}^{(ij)})> \Omega$, we `lose'. 
If $\hat{t}^{(ij)} > \tau_{\rm max}$, go to 1. 
\end{itemize}
\end{enumerate}
\end{enumerate}
Then, we repeat the above procedure for all possible pairs of  $1,784$ stocks 
listed on the first section of the Tokyo Stock Exchange leading up to totally ${}_{1,784}C_{2}=1,590,436$ pairs. 
We play our game according to 
the above algorithm for each pair, and if a pair 
$(i,j)$ passes their decision time $\hat{t}^{(ij)}$ 
resulting in 
the profit: 
\begin{equation}
d_{ij}(t_{<}^{(ij)})-d_{ij}(\hat{t}^{(ij)}) \,\,\,\mbox{with}\,\,\,
\hat{t}^{(ij)}=t_{>}^{(ij)}
\end{equation}
or the loss: 
\begin{equation}
d_{ij}(\hat{t}^{(ij)})-d_{ij}(t_{<}^{(ij)})\,\,\,\mbox{with}\,\,\, 
\hat{t}^{(ij)}=t_{*}^{(ij)},
\end{equation}
we discard the pair $(i,j)$ and 
never `recycle' the pair again for pairs trading. 
Of course, such treatment might be hardly accepted in realistic 
pairs trading because traders tend to 
use the same pairs as the one which gave them a profit in the past markets. 
Nevertheless, here we shall utilize this somewhat `artificial' treatment in order to 
quantify the performance of paris trading through the measurements 
$p_{w}$ and $\eta$ systematically. 
We also simplify the game by restricting ourselves to 
the case in which each trader always makes a trade by a unit volume.

In the next section, we show several result of empirical data analysis.  
\section{Empirical data analysis}
\label{sec:sec4}
Here we show several empirical data analyses done for all possible pairs of  $1,784$ stocks 
listed on the first section of the Tokyo Stock Exchange leading up to ${}_{1,784}C_{2}=1,590,436$ pairs. 
The daily data sets are collected for the past four years 2009-2010 from 
the web cite \cite{Yahoo}. 
In our empirical analysis, we set 
$\tau=250$ [days], 
$\rho_{0}=0.6, \sigma_{\rm min}=0.05, \sigma_{\rm max}=0.2$. 
\subsection{Preliminary experiments}
Before we show our main result, we provide the two empirical distributions for 
the correlation coefficients and volatilities, which might posses very 
useful information about selecting the active pairs. 
We also discuss the distribution of 
the first-passage time to quantify the processing time roughly. 
\subsubsection{Correlation coefficients and volatilities}
\begin{figure}[htb]
\begin{center}
\includegraphics[width=5.8cm]{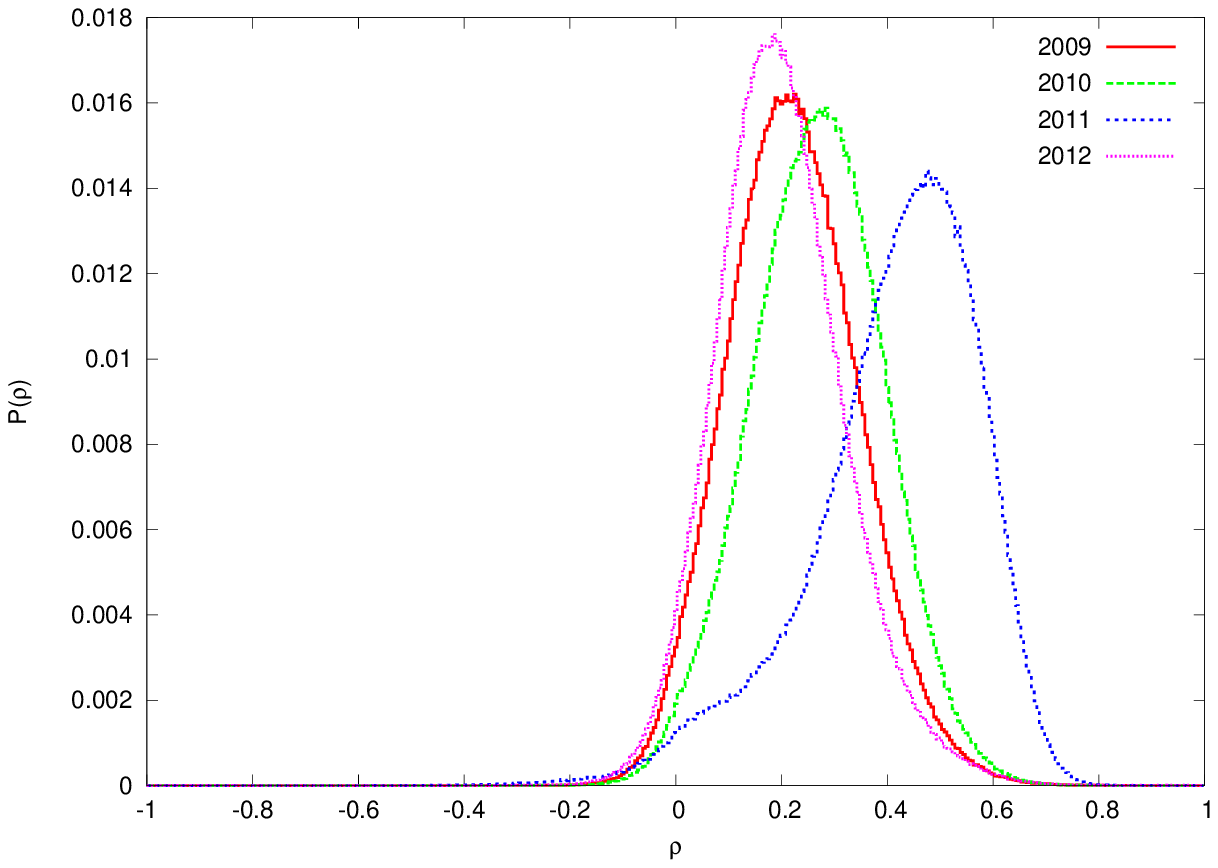}
\includegraphics[width=5.8cm]{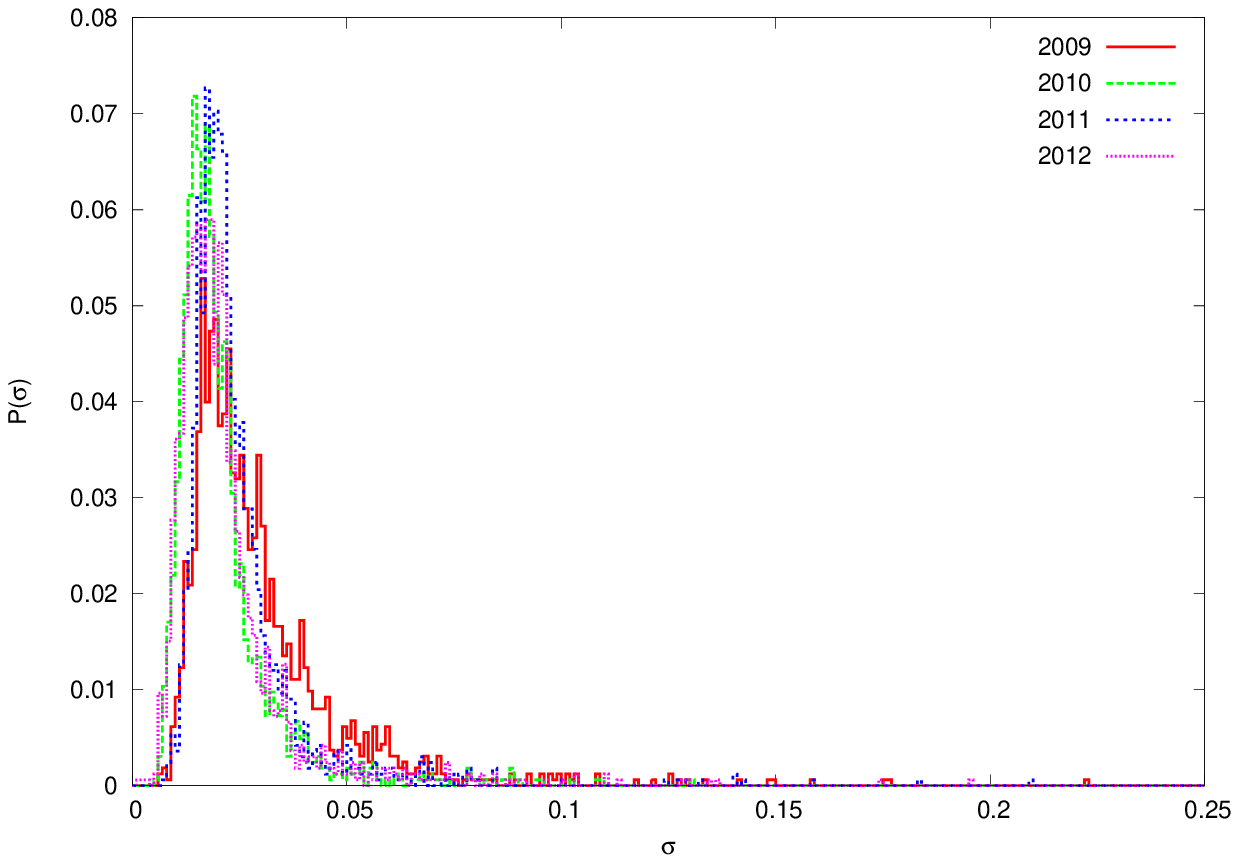}
\caption{\footnotesize 
Distributions  of Pearson estimator $\{\rho_{ij}(t)\}$ (left) and 
volatility $\{\sigma_{i}(t)\}$ (right) for 
the past four years data sets:  2009, 2010, 2011 and 2012. 
On the other hand, the distribution of the volatility 
is almost independent of year and it possess a single peak around $0.02$. 
}
\label{fig:cor}
\end{center}
\end{figure}

In Fig. \ref{fig:cor}, 
we plot the distributions of 
$\{\rho_{ij}(t)\}$ (left) and 
$\{\sigma_{i}(t)\}$ (right) for 
the past four years (2009-2012).  
From the left panel, we find that 
the distribution of correlation coefficients 
is apparently skewed for all years and 
the degree of skewness in 2011 is the 
highest among the four due to 
the great east Japan earthquake as reported in \cite{Ibuki,Ibuki2,Ibuki3}. 
Actually, we might observe that most of stocks in the multidimensional scaling plane 
shrink to a finite restricted region due to the strong correlations. 

On the other hand, the distribution of the volatility 
is almost independent of the year and possess a peak around $0.02$. 
We are 
confirmed from these empirical distributions  
that the choice of 
the system parameters 
$\rho_{0}=0.6, \sigma_{\rm min}=0.05, \sigma_{\rm max}=0.2$ 
could be justified properly in the sense that 
the number of pairs $(i,j)$ satisfying 
the criteria $\rho_{ij}(t) <\rho_{0}$ and 
$\sigma_{\rm min} <\sigma_{i}(t) < \sigma_{\rm max}$ 
is not a vanishingly small fraction but 
reasonable number of pairs ($\sim 300$) can remain in the system. 
\subsubsection{First-passage times}
\begin{figure}[ht]
\begin{center}
\includegraphics[width=5.8cm]{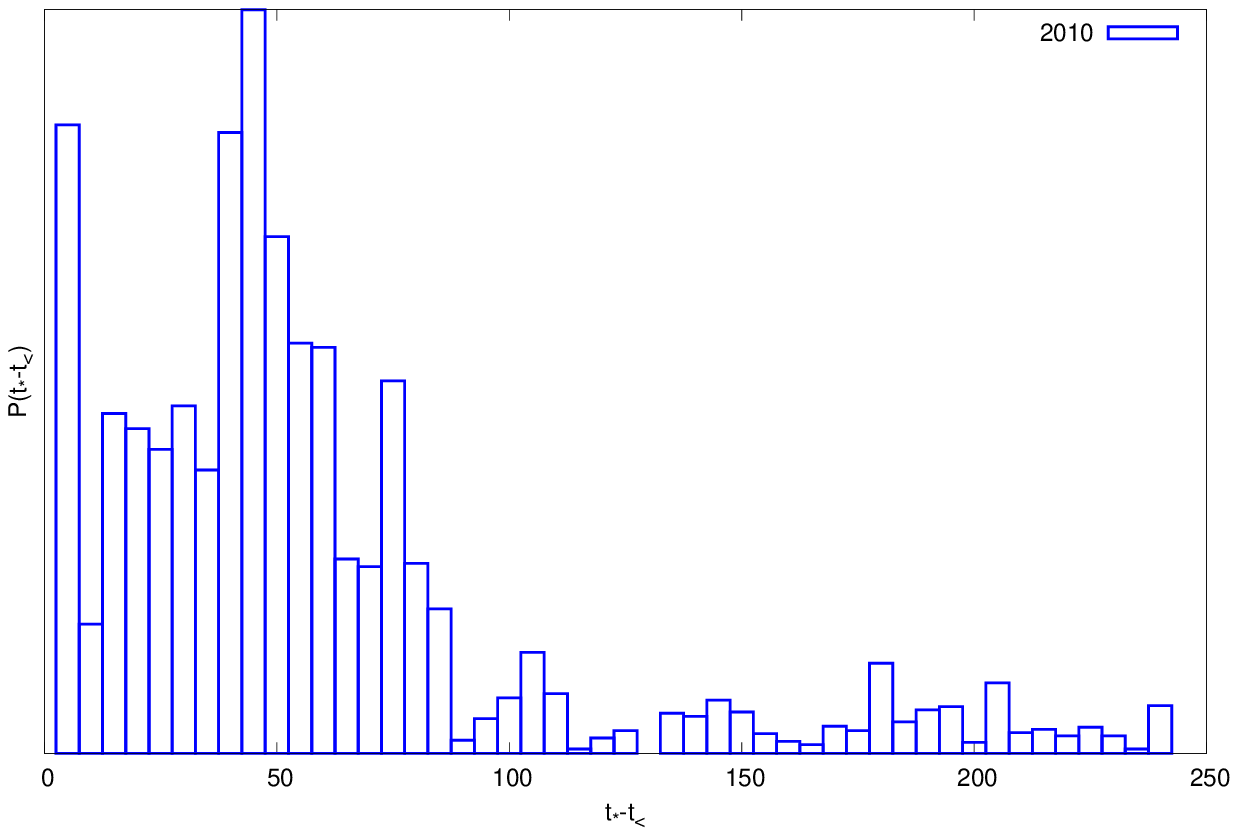}
\includegraphics[width=5.8cm]{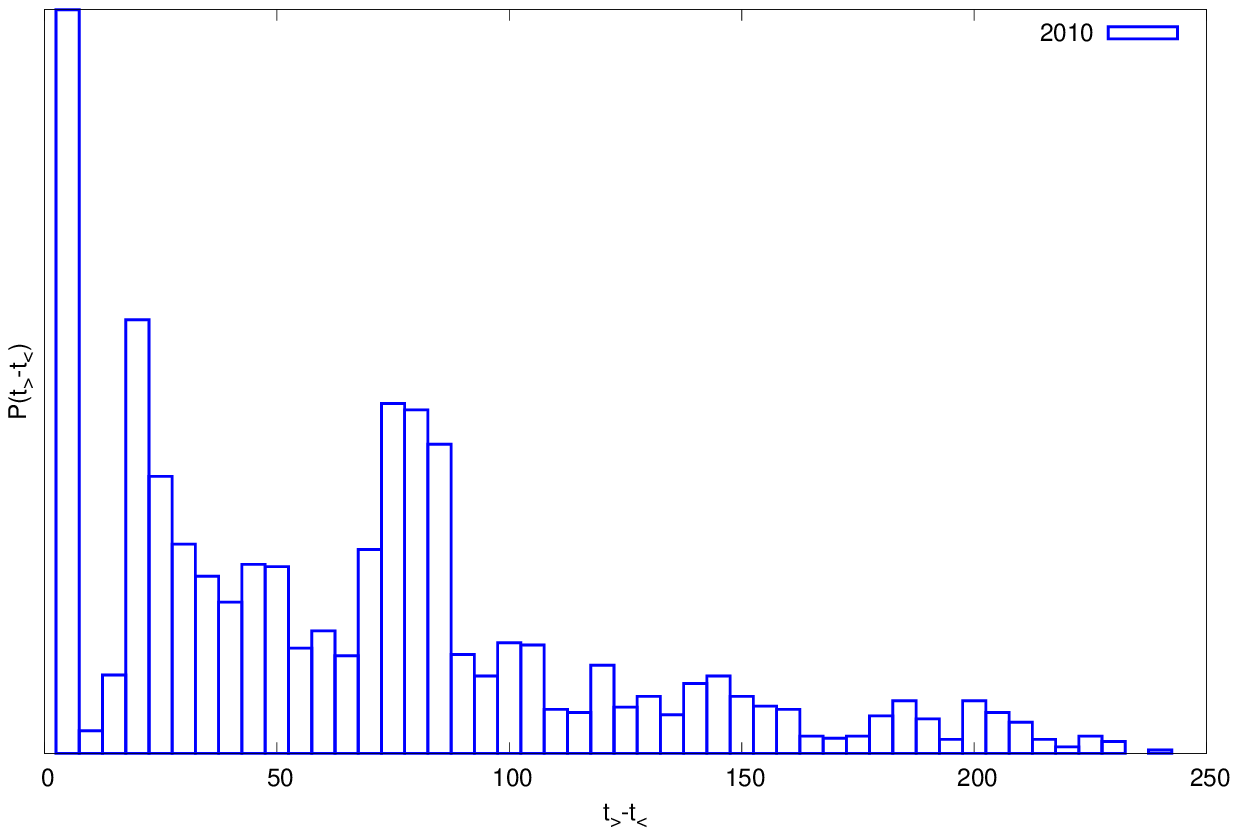}
\end{center}
\caption{\footnotesize 
The empirical distributions $P(t)$ of first-passage times  
$t \equiv \hat{t}-t_{<}$. 
The left panel is $P(t \equiv t_{*}-t_{<})$, whereas the right is 
$P(t \equiv t_{>}-t_{<})$. 
We find that one confirms the lose 
by loss-cutting by 50 days after the 
start point $t_{<}$ in most cases and 
the decision is disclosed at latest by 250 days after 
the $t_{<}$. 
On the other hand, 
we win within several days after the start and 
a single peak is actually located in the short time frame. 
}
\label{fig:fg0}
\end{figure}

We next show the distributions of the first-passage times 
for the data set in 2010. 
It should be noted that 
we observe the duration $t$ as a first passage time 
from the point $t_{<}$ in time axis, hence, 
the distributions of the duration $t$ are 
given for 
\begin{eqnarray}
P(t) & = & P(t \equiv  t_{>}-t_{<}) \,\,\,\,\,\mbox{(for win)}, \\
P(t) & = & P(t \equiv  t_{*}-t_{<})\,\,\,\,\,\,\mbox{(for lose)},
\end{eqnarray}
respectively. 
We plot the results in Fig. \ref{fig:fg0}. 
From the left panel, 
we find that one confirms the lose 
by loss-cutting by 50 days after the 
start point $t_{<}$ in most cases,  and 
the decision is disclosed at latest by 250 days after 
the $t_{<}$. 
On the other hand, 
we win within several days after the start and 
a single peak is actually located in the short time frame. 
These empirical findings 
tell us that in most cases, 
the spread between highly-correlated two stocks 
actually shrink shortly even if the two prices of 
the stocks exhibit `mis-pricing' leading up to a large spread temporally. 
Taking into account this fact, 
our findings also imply that 
the selection by correlation coefficients and volatilities 
works effectively to make the pairs trading useful. 
\subsection{Winning probability}
\begin{figure}[ht]
\begin{center}
\includegraphics[width=5.8cm]{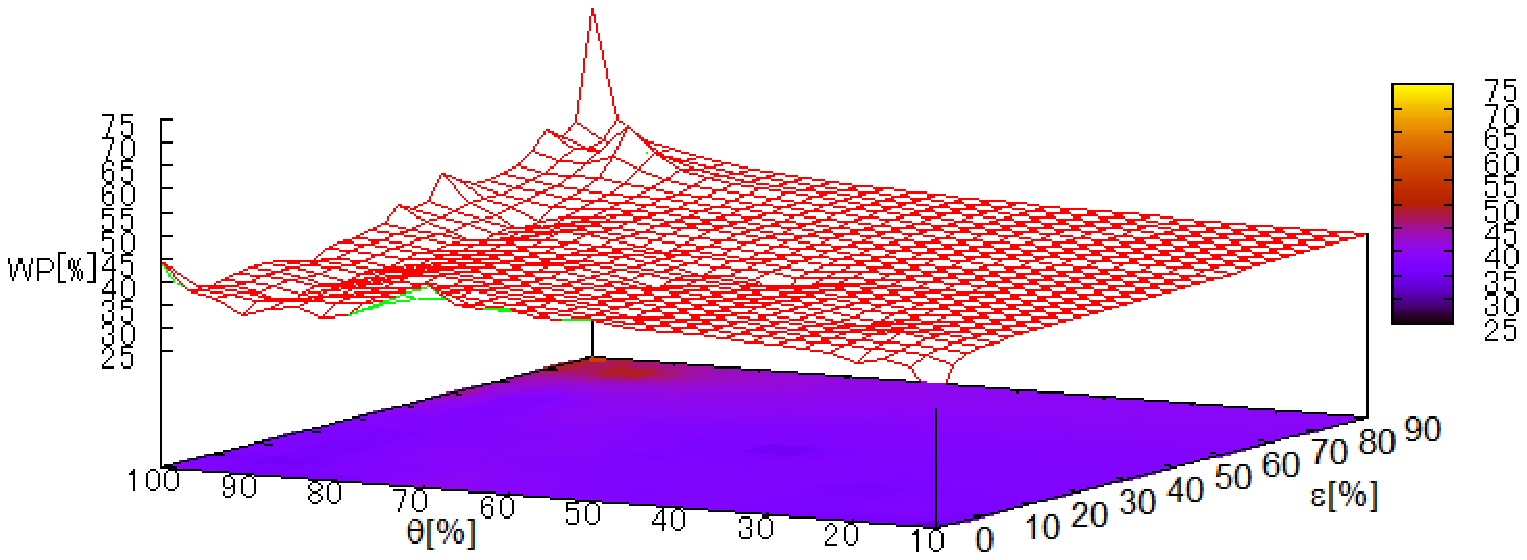} 
\includegraphics[width=5.8cm]{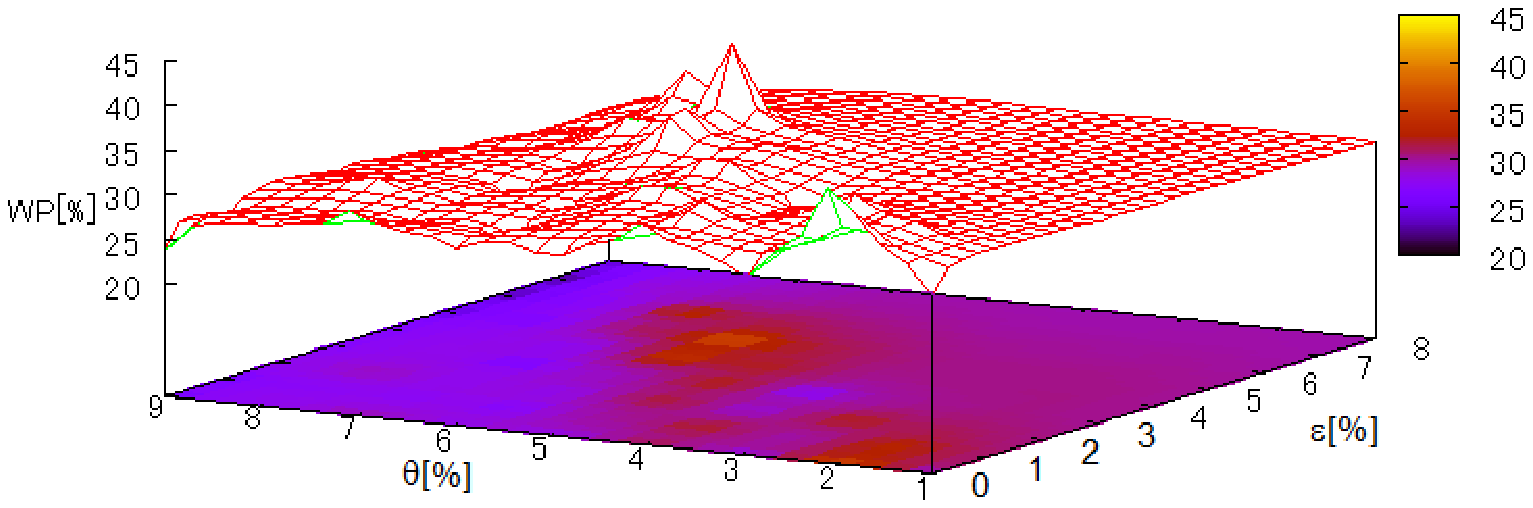}
\includegraphics[width=5.8cm]{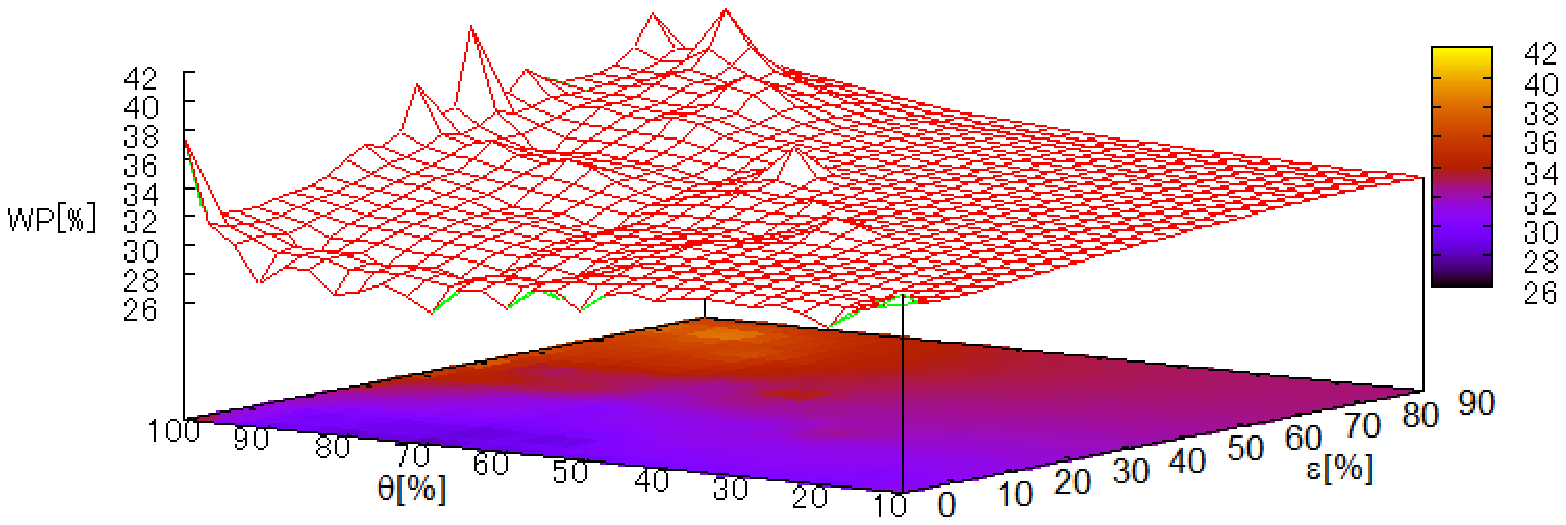}
\includegraphics[width=5.8cm]{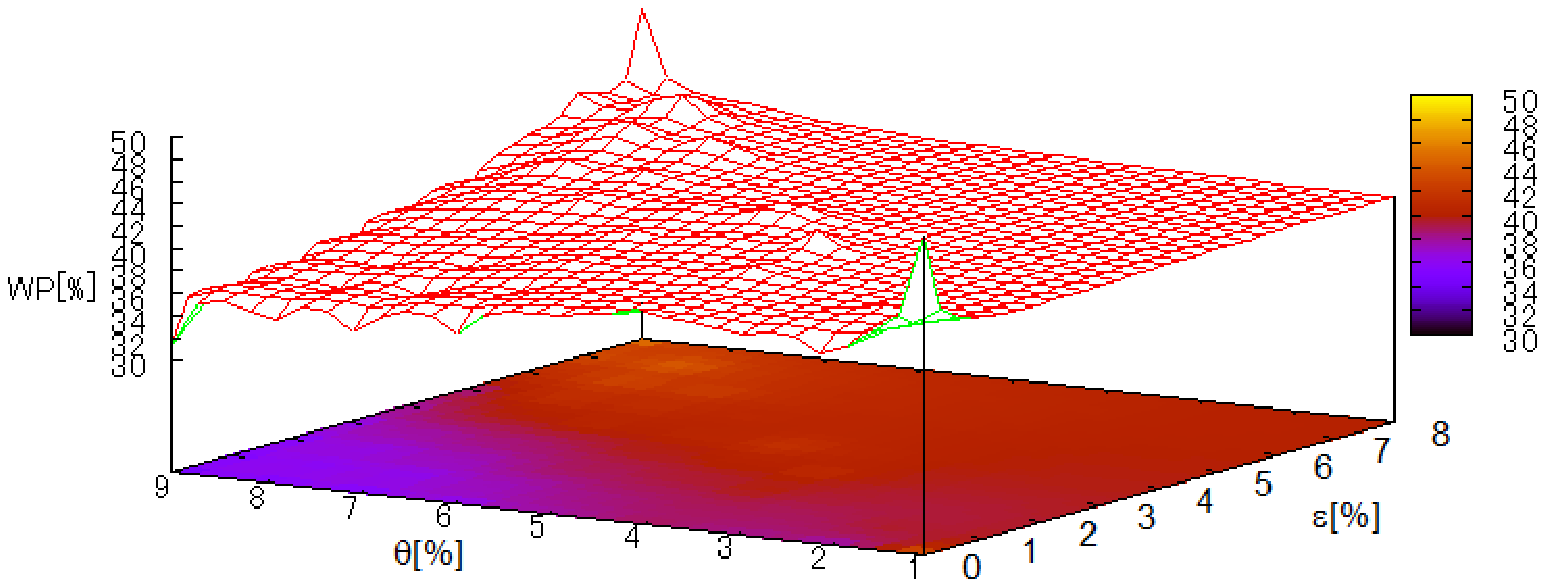}
\includegraphics[width=5.8cm]{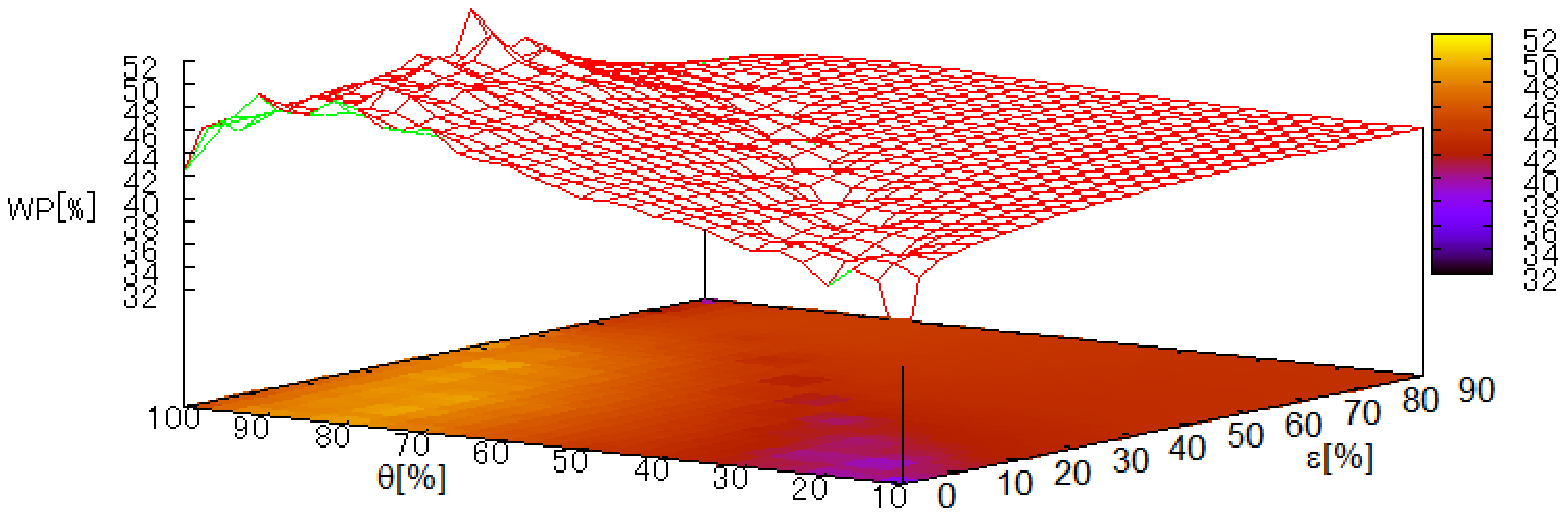}
\includegraphics[width=5.8cm]{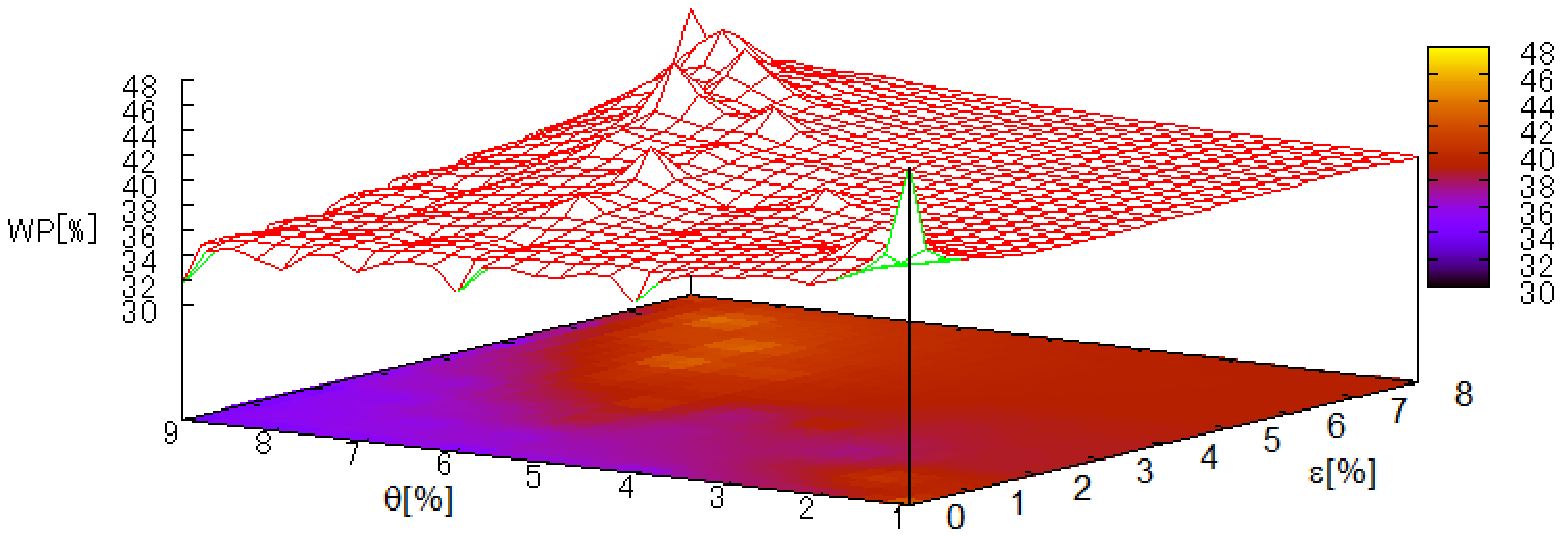}
\end{center}
\caption{\footnotesize 
Winning probability $p_{w}$ as a function of 
$(\theta,\varepsilon)$. From the top most to the bottom, the results in 
2012, 2011 and 2010 are plotted. 
The right panels are the plots for relatively small range of thresholds $(\theta,\varepsilon)$. 
From these panels, we find that  
the winning probability is less than that of the `draw case' $p_{w}=0.5$ in most cases of the 
thresholds $(\theta,\varepsilon)$. 
See also Table \ref{tab:tb1} (2012) and Table \ref{tab:tb2} (2011) for the details. 
}
\label{fig:fg1}
\end{figure}

As our main results, 
we first show the wining probability $p_{w}$ as a function of thresholds $(\theta,\varepsilon)$ 
defined by (\ref{eq:pw}) in Fig. \ref{fig:fg1}. 
To show it effectively,  we display the results as three dimensional plots with contours. 
From these panels, we find that 
the winning probability is unfortunately 
less than that of the `draw case' $p_{w}=0.5$ in most choices of the 
thresholds $(\theta,\varepsilon)$. 
We also find that 
for a given $\theta (>\varepsilon)$, 
the probability $p_{w}$ is almost a monotonically 
increasing function of  $\theta$ in all the three years.  
This result is naturally accepted because the trader might take more careful 
actions on the starting of the pairs trading for a relatively large $\theta$. 
\mbox{} 

To see the result more carefully, we 
write the raw data produced by our analysis 
in Table \ref{tab:tb1} (2012) and Table \ref{tab:tb2} (2011). 
From these two tables, we find that 
relatively higher 
winning probabilities  
$p_{w} \sim 0.7$ are observed, however, 
for those cases, 
the number of wins $N_{w}$ (or loses $N_{l}$) is small, 
and it should be more careful for us to evaluate the winning possibility of pairs trading from 
those limited data sets.  
\subsection{Profit rate}
\begin{figure}[ht]
\begin{center}
\includegraphics[width=5.8cm]{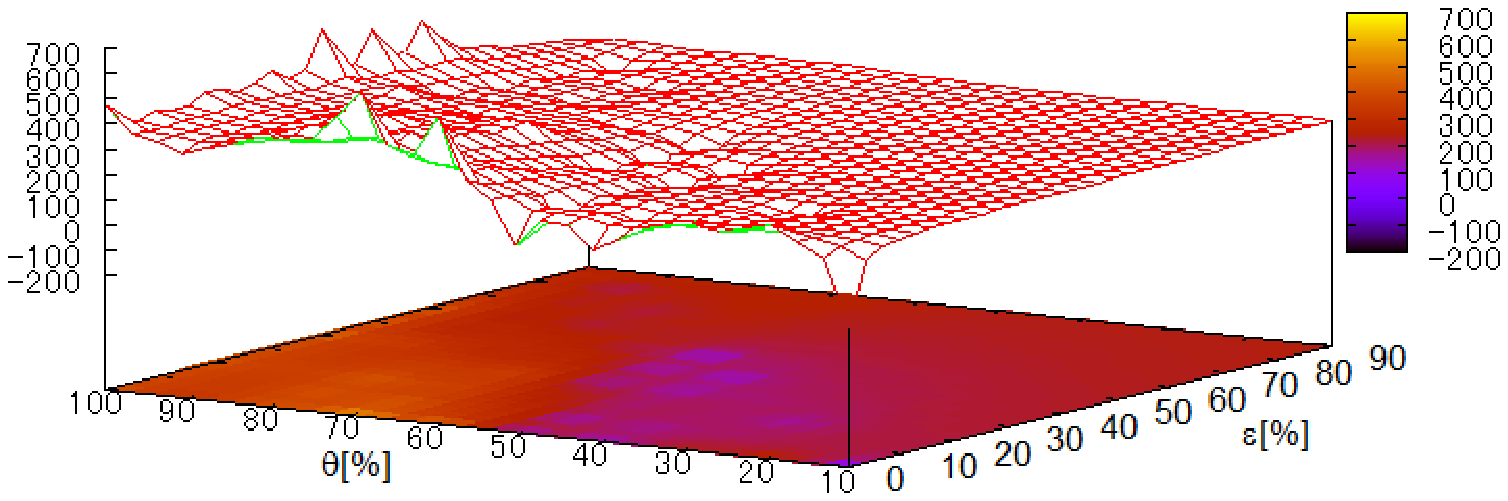}
\includegraphics[width=5.8cm]{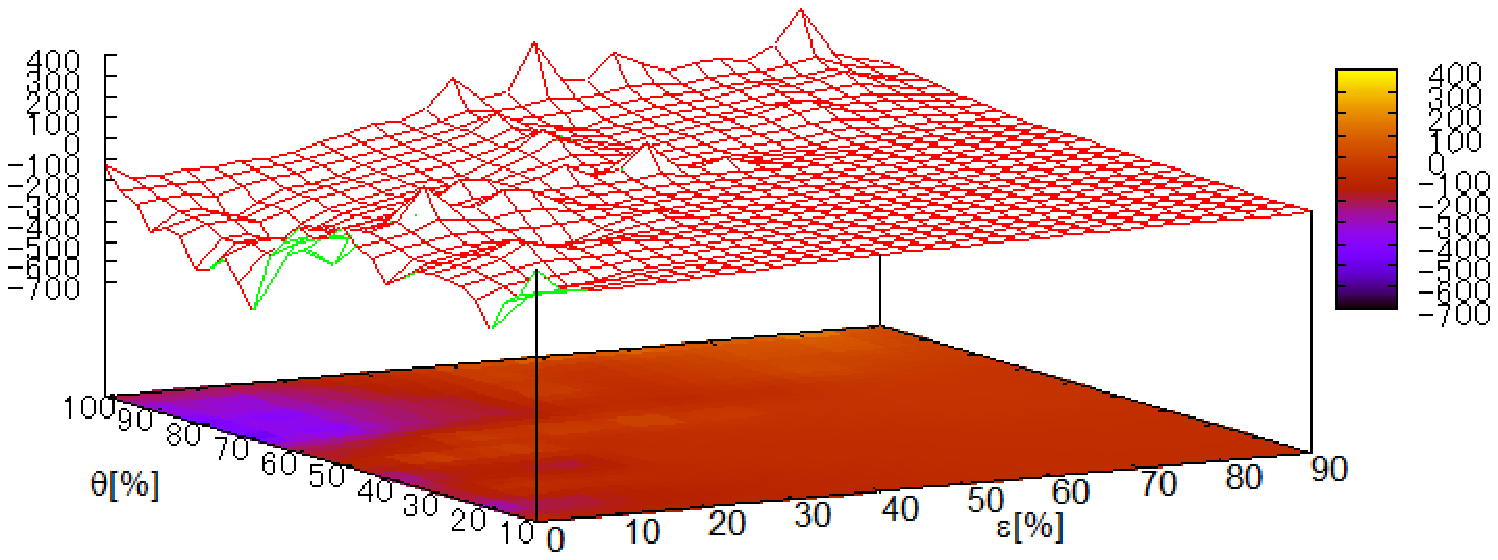} 
\includegraphics[width=5.8cm]{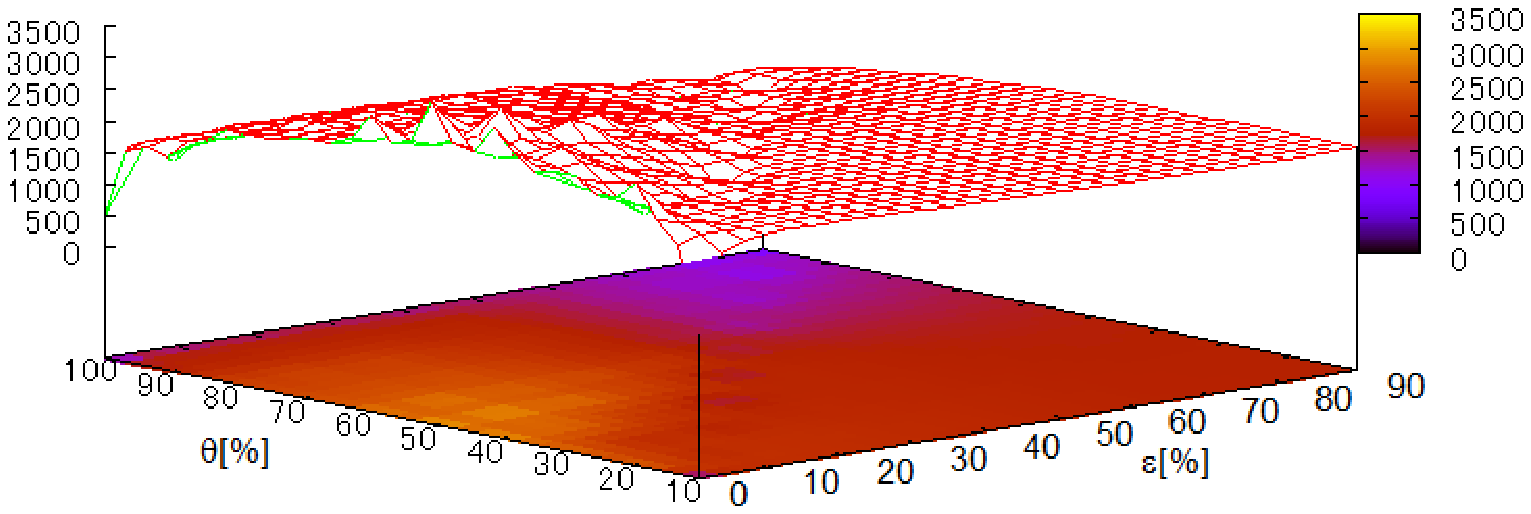} 
\caption{\footnotesize 
Profit rate $\eta$ as a function of 
$(\theta,\varepsilon)$. From the upper left  to the bottom, the results in 
2012, 2011 and 2010 are plotted. 
We clearly find that for almost all of the combinations $(\theta,\varepsilon)$, 
one can obtain the positive profit rate $\eta>0$, which means that our algorithm actually achieves 
almost risk-free asset management and it might be a justification of the usefulness of pairs trading.
}
\label{fig:fg_gain2}
\end{center}
\end{figure}

In order to consider the result obtained by our algorithm for pairs trading 
from a slightly different aspect, 
we plot the profit rate $\eta$ given by (\ref{eq:profit}) as a function 
of thresholds $(\theta,\varepsilon)$ in Fig. \ref{fig:fg_gain2}. 
We clearly find that for almost all of the combinations $(\theta,\varepsilon)$, 
one can obtain the positive profit rate $\eta>0$, which means that our algorithm actually achieves 
almost risk-free asset management and it might be a justification of the usefulness of pairs trading. 

At a glance, it seems 
that the result of the small winning 
probability $p_{w}$ is inconsistent with that of 
the positive profit rate $\eta>0$. However, 
the result can be possible to be obtained. 
To see it explicitly, 
let us assume that the pairs 
$(i,j)$ and $(k,l)$ lose and 
the pair $(m,n)$ wins for 
a specific choice of thresholds $(\theta_{+},\varepsilon_{+})$. 
Then, the wining probability is $p_{w}=1/3$. 
However, 
the profits for these three pairs could satisfy the following inequality: 
\begin{equation}
d_{mn}(t_{>}^{(mn)})-d_{mn}(t_{>}^{(mn)}) > 
\{d_{ij}(t_{>}^{(ij)})-d_{ij}(t_{*}^{(ij)})\}
+
\{d_{kl}(t_{>}^{(kl)})-d_{kl}(t_{*}^{(kl)})\}
\end{equation}
From the definition of the profit rate (\ref{eq:profit}),
we are immediately conformed as 
\begin{eqnarray}
\eta (\theta_{+},\varepsilon_{+}) & = &  
d_{mn}(t_{>}^{(mn)})-d_{mn}(t_{>}^{(mn)}) -
\{d_{ij}(t_{>}^{(ij)})-d_{ij}(t_{*}^{(ij)})\} \nonumber \\
\mbox{} & - & 
\{d_{kl}(t_{>}^{(kl)})-d_{kl}(t_{*}^{(kl)})\} >  0. 
\end{eqnarray} 
Hence, 
an active pair producing a relatively large arbitrage 
can compensate 
the loss of wrong active pairs by choosing 
the threshold $(\theta,\varepsilon)$ appropriately.  
It might be an ideal scenario for the pairs trading.  
\mbox{}

Finally, we should stress that 
the fact $\eta > 0$ in most cases of 
thresholds $(\theta,\varepsilon)$ 
implies that 
automatic pairs trading system could be constructed 
by applying our algorithm for all possible $(\theta,\varepsilon)$ in parallel. 
However, it does not mean 
that we can always obtain positive profit `actually'.  
Our original motivation in this paper is just to examine (from the stochastic properties
of spreads between two stocks) how much percentage of highly correlated pairs
is suitable for the candidate in pairs trading in a specific market, namely, Tokyo Stock Exchange.
In this sense, our result could not be used directly for practical trading.
Nevertheless, as one can easily point out, we may pare down the candidates by introducing
the additional transaction cost, and even for such a case,  the game 
to calculate the winning probability etc. by regarding the trading
as a mixture of first-passage processes might be useful.
\subsection{Profit rate versus volatilities}
\begin{figure}[ht]
\begin{center}
\includegraphics[width=8cm]{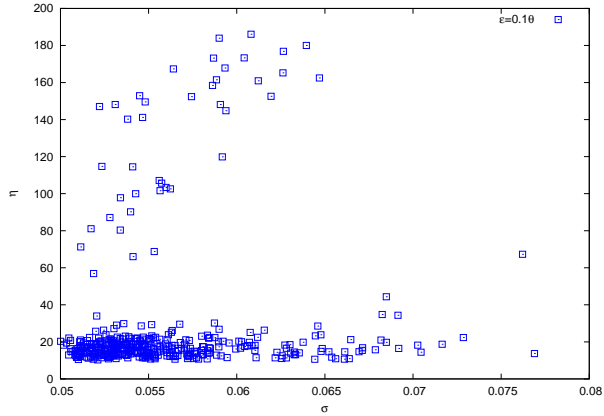}
\end{center}
\caption{\footnotesize 
The profit rate $\eta$ against the volatilities $\sigma$ as a scattergram. 
We set the profit-taking threshold $\varepsilon$ as 
$\varepsilon = 0.1 \theta$
and vary the starting threshold $\theta$ in the range of $0.1 \leq \theta \leq  0.3$. 
We find that 
there exist two distinct clusters (components) in the winner pairs, 
namely, 
the winner pairs giving us the profit rate typically as much as 
$\eta \simeq \theta -\varepsilon =\theta -0.1\theta =0.9\, \theta \simeq 0.2$ 
for the range of $0.1 \leq \theta \leq  0.3$, which are almost independent of $\sigma$,
and the winner pairs having the profit linearly dependent on the volatility $\sigma$. 
Note that the vertical axis is shown as `percentage'. 
}
\label{fig:fg_scat}
\end{figure}

In Fig. \ref{fig:fg_scat}, 
we plot the profit rate $\eta$ against the volatilities $\sigma$ as a scattergram only for the winner pairs. 
In this plot, we set the profit-taking threshold $\varepsilon$ as 
\begin{equation}
\varepsilon = 0.1 \theta
\end{equation}
and vary the starting threshold $\theta$ in the range of $0.1 <\theta < 0.3$. 
For each active winner pair, 
we observe the profit rate $\eta$ and the average volatility $\sigma$ of the two stocks in each pair, 
and plot the set $(\sigma,\eta)$ in the two-dimensional scattergram. 
From this figure, we find that 
there exist two distinct clusters (components) in the winner pairs, 
namely, 
the winner pairs giving us the profit rate typically as much as 
\begin{equation}
\eta \simeq \theta -\varepsilon =\theta -0.1\theta =0.9\, \theta \simeq 0.2
\label{eq:eta02}
\end{equation}
for the range of $0.1  \leq \theta \leq 0.3$, which are almost independent of $\sigma$,
and the winner pairs having the profit rate linearly dependent on the volatility $\sigma$. 
The former is a low-risk group, whereas the latter is a high-risk group. 
The density of the points for the low-risk group in Fig. \ref{fig:fg_scat} is much higher than that of the high-risk group. 
Hence, we are confirmed that our selection procedure of the active pairs 
works effectively to manage the asset as safely as possible by reducing the risk 
which usually increases as the volatility grows.  

Finally, it should be noted that as we discussed in subsection \ref{sec:sub_constraint}, 
the value $\theta-\varepsilon$ is a lower bound of the 
profit rate (see equations (\ref{eq:lower_bound}) and (\ref{eq:eta02})). 
Therefore, in the above case, the lower bound for the profit rate $\eta$ should be estimated for 
$0.1 \leq  \theta \leq 0.3$ as 
\begin{equation}
\eta \geq \eta_{\rm min}=0.9\, \theta =0.9 \times 0.1 = 0.09.
\label{eq:lower_bound2}
\end{equation}
The lowest value for the profit rate 
(\ref{eq:lower_bound2}) 
is consistent with 
the actually observed lowest value in 
the scattergram shown in Fig. \ref{fig:fg_scat}. 
\subsection{Examples of winner pairs}
Finally, we shall list several examples of active pairs to win the game. 
Of course, we cannot list all of the winner pairs in this paper, hence, 
we here list only three pairs as examples, 
each of which includes 
{\it SANYO SPECIAL STEEL Co. Ltd.} (ID: {\tt 5481}) 
and the corresponding partners are 
{\it HITACHI METALS. Ltd.} (ID: {\tt 5486}), 
{\it MITSUI MINING $\&$ SMELTING Co. Ltd.} (ID: {\tt 5706}) and 
{\it PACIFIC METALS Co. Ltd.} (ID: {\tt 5541}). 
Namely, the following three pairs 
\[
({\tt 5481},{\tt 5486}),
({\tt 5481},{\tt 5706}),({\tt 5481},{\tt 5541})
\]
actually won in our empirical analysis of the game. 
Note that each ID in the above expression corresponds to each identifier used in Yahoo!Finance \cite{Yahoo}. 
As we expected as an example of 
{\it coca cola} and {\it pepsi cola} \cite{Pairs}, these are all the same type of industry 
(the steel industry). 
We would like to stress that 
we should act with caution 
to trade using the above pairs because 
the pairs just won the game in which 
the pairs once got a profit  in the past are never recycled in future. 
Therefore, we need much more extensive analysis for the above pairs 
to use them in practice. 
\section{Summary}
In this paper, 
we proposed a very simple and effective algorithm to make the pairs trading easily and automatically. 
We applied our algorithm to daily data of stocks in the first section of the Tokyo Stock Exchange.  
Numerical evaluations of the algorithm for the empirical data set 
were carried out for all possible pairs 
by changing the starting ($\theta$), profit-taking ($\varepsilon$) and 
stop-loss ($\Omega$) conditions 
in order to look for the best possible combination of the conditions $(\theta,\varepsilon,\Omega)$. 
We found that for almost all of the combinations $(\theta,\varepsilon)$  under the constraint $\Omega=2\theta - \varepsilon$, 
one can obtain the positive profit rate $\eta>0$, which means that our algorithm actually achieves 
almost risk-free asset management at least for the past three years (2010-2012) 
and it might be a justification of the usefulness of pairs trading. 
Finally, we showed several examples of active pairs to win the game. 
As we expected before, the pairs are all the same type of industry (for these examples, it is the steel industry). 
We should conclude that the fact $\eta > 0$ in most cases of 
thresholds $(\theta,\varepsilon)$ 
implies that automatic pairs trading system could be constructed 
by applying our algorithm for all possible $(\theta,\varepsilon)$ in parallel way. 

Of course, the result does not mean directly 
that we can always obtain positive profit in a practical pairs trading.  
Our aim in this paper was to examine how much percentage of highly correlated pairs
is suitable for the candidate in pairs trading in a specific market, namely, Tokyo Stock Exchange.
In this sense, our result could not be used directly for practical pairs trading.
Nevertheless, we may pare down the candidates by introducing
the additional transaction cost, and even for such a case,  the game 
to calculate the winning probability etc. by regarding the trading
as a mixture of first-passage processes might be useful.

We are planning to consider pairs listed in different stock markets, 
for instance, one is in Tokyo and the other is in NY.
Then, of course, we should also consider the effect of the exchange rate. 
Those analyses  might be addressed as our future study. 
\begin{table}[ht]
\begin{center}
\begin{tabular}{rrr|rrr||rrr|rrr}
\hline
$\varepsilon$[\%] & $\theta$[\%] & $\Omega$[\%] & $N_{w}$ & $N_{l}$  & $p_{w}$ [\%] & $\varepsilon$[\%] & $\theta$[\%] & $\Omega$[\%] & $N_{w}$
 & $N_{l}$ & $p_{w}$[\%]\\ \hline
0&10&20&11&32&25.6&0&80&160&5&13&27.7 \\
0&20&40&25&52&32.4&10&80&150&6&9&40.0 \\
10&20&30&23&33&41.0&20&80&149&8&9&47.0 \\
0&30&60&24&42&36.3&30&80&130&8&12&40.0 \\
10&30&50&23&41&35.9&40&80&120&9&14&39.1 \\
20&30&40&17&41&29.3&50&80&110&10&12&45.4\\
0&40&80&22&36&37.9&60&80&100&9&12&42.8 \\
10&40&70&22&35&38.5&70&80&90&8&11&42.1 \\
20&40&60&22&32&40.7&0&90&180&4&9&30.7 \\
30&40&50&17&27&38.6&10&90&170&4&12&25.0 \\
0&50&100&15&23&39.4&20&90&160&4&11&26.6 \\
10&50&90&16&31&34.0&30&90&150&4&9&30.7 \\
20&50&80&18&30&37.5&40&90&140&5&9&35.7 \\
30&50&70&16&27&37.2&50&90&130&4&10&28.5 \\
40&50&60&9&25&26.4&60&90&120&3&10&23.0 \\
0&60&120&11&16&40.7&70&90&110&4&6&40.0 \\
10&60&110&14&20&41.1&80&90&100&5&2&71.4 \\
20&60&100&13&22&37.0&0&100&200&4&5&44.4 \\
30&60&90&13&28&31.7&10&100&190&4&8&33.3 \\
40&60&80&14&26&35.0&20&100&180&4&6&40.0 \\
50&60&70&10&17&37.0&30&100&170&4&9&30.7 \\
0&70&140&11&10&52.3&40&100&160&5&8&38.4\\
10&70&130&11&14&44.0&50&100&150&5&6&45.4 \\
20&70&120&10&16&38.4&60&100&140&7&5&58.3 \\
30&70&110&13&16&44.8&70&100&130&5&6&45.4 \\
40&70&100&13&19&40.6&80&100&120&5&4&55.5 \\
50&70&90&11&21&34.3&90&100&110&3&1&75.0 \\
60&70&80&13&18&41.9&&&&& \\
\hline
\end{tabular}
\end{center}
\caption{\footnotesize 
Details of the result in 2012. See also Fig. \ref{fig:fg1} (the top most). 
We find that 
relatively higher 
winning probabilities  
$p_{w} \sim 0.7$ are observed, however, 
the number of wins $N_{w}$ (and lose $N_{l}$) is small. 
It should be noted that 
$\Omega = 2\theta -\varepsilon$ holds. 
}
\label{tab:tb1}
\end{table}
\begin{table}[ht]
\begin{center}
\begin{tabular}{rrr|rrr||rrr|rrr}
\hline
$\varepsilon$[\%] & $\theta$[\%] & $\Omega$[\%] & $N_{w}$ & $N_{l}$  & $p_{w}$ [\%] & $\varepsilon$[\%] & $\theta$[\%] & $\Omega$[\%] & $N_{w}$
 & $N_{l}$ & $p_{w}$[\%]\\ \hline
0&10&20&93&200&31.7&0&80&160&9&30&23.0\\
0&20&40&93&245&27.5&10&80&150&11&34&24.4 \\
10&20&50&91&221&29.1&20&80&149&17&39&30.3 \\
0&30&60&72&178&28.8&30&80&130&16&42&27.5 \\
10&30&50&90&198&31.2&40&80&120&25&42&37.3 \\
20&30&40&87&195&30.8&50&80&110&25&54&31.6\\
0&40&80&44&111&28.3&60&80&100&34&53&39.0 \\
10&40&70&61&135&31.1&70&80&90&41&49&45.5 \\
20&40&60&69&162&29.8&0&90&180&6&18&25.0 \\
30&40&50&87&147&37.1&10&90&170&7&22&24.1 \\
0&50&100&31&82&27.4&20&90&160&11&25&30.5 \\
10&50&90&41&88&31.7&30&90&150&13&26&33.3 \\
20&50&80&50&103&32.6&40&90&140&18&31&36.7 \\
30&50&70&50&123&28.9&50&90&130&15&38&28.3 \\
40&50&60&57&114&33.3&60&90&120&23&39&37.0 \\
0&60&120&21&57&26.9&70&90&110&28&44&38.8 \\
10&60&110&21&69&23.3&80&90&100&31&33&48.4 \\
20&60&100&32&74&30.1&0&100&200&6&10&37.5 \\
30&60&90&34&80&29.8&10&100&190&6&13&31.5 \\
40&60&80&45&89&33.5&20&100&180&8&17&32.0\\
50&60&70&53&86&38.1&30&100&170&11&19&36.6 \\
0&70&140&14&47&22.9&40&100&160&14&21&40.0\\
10&70&130&14&48&22.5&50&100&150&17&21&44.7 \\
20&70&120&21&53&28.3&60&100&140&18&26&40.9 \\
30&70&110&22&64&25.5&70&100&130&19&33&36.5 \\
40&70&100&29&67&30.2&80&100&120&23&31&42.5 \\
50&70&90&32&75&29.9&90&100&110&20&34&37.0 \\
60&70&80&38&70&35.1&&&&&& \\
\hline
\end{tabular}
\end{center}
\caption{\footnotesize 
Details of the result in 2011. See also Fig. \ref{fig:fg1} (the middle). 
It should be noted that 
$\Omega = 2\theta -\varepsilon$ holds. 
}
\label{tab:tb2}
\end{table}
\subsubsection*{Acknowledgements}
This work was financially supported by Grant-in-Aid for Scientific Research (C) of
Japan Society for the Promotion of Science No. 2533027803 and 
Grant-in-Aid for Scientific Research (B) of
Japan Society for the Promotion of Science No. 26282089. 
We were also supported by 
Grant-in-Aid for 
Scientific Research on Innovative Area No. 2512001313. 
One of the authors (JI) thanks 
Anirban Chakraborti for his useful comments on this study at the early stage. 
\subsection*{Conflict of interest statement}
On behalf of all authors, the corresponding author (JI) states that there is no conflict of interest.

\end{document}